\documentclass[aps,twocolumn,nofootinbib,prl]{revtex4}
\usepackage{hyperref}
\usepackage{graphicx}
\usepackage{color}
\def\gr{$\gamma$-ray}

\begin{document}
\title{Low-energy break in the spectrum of Galactic cosmic rays.}
\author{A.Neronov$^1$, D.V.Semikoz$^{2,3}$, A.M.Taylor$^1$}
\affiliation{$^1$ ISDC Data Center for Astrophysics, Ch.d'Ecogia 16, 1290, Versoix, Switzerland\\
$^2$ APC, 10 rue Alice Domon et Leonie Duquet, F-75205 Paris Cedex 13, France }
\begin{abstract}
Measurements of the low energy spectrum of Galactic Cosmic Rays (GCR) by detectors at or near the Earth are affected by Solar modulation. To overcome this difficulty, we consider nearby molecular clouds as GCR detectors outside the Solar system.
Using \gr\ observations of the clouds by the Fermi telescope we derive the spectrum of GCRs in the clouds from the observed \gr\ emission spectrum. We find  that the GCR spectrum has a low energy break with the spectral slope hardening by $\Delta \Gamma= 1.1\pm 0.3$ at an energy of  $E=9\pm 3$~GeV.  Detection of a low-energy break enables a measurement of GCR energy density in the interstellar space $U=0.9\pm 0.3$~eV/cm$^3$.
\end{abstract}

\maketitle

\noindent{\bf Introduction.}
The spectrum of cosmic rays (CR) with energies  $E< 100$~GeV measured by CR detectors inside the Solar System  \cite{bess,pamela,webber} is not identical to the Galactic CR (GCR) spectrum due to the extinction of the low-energy CRs by the Solar wind \cite{parker,gloekler67,florinski}.  Uncertainties in the knowledge of  properties of the Solar wind, its termination shock, Heliosheath and Heliotail   introduce uncertainties in our knowledge of GCR spectrum. 
 
The only possibility to measure the GCR spectrum unaffected by the Solar modulation is to consider nearby mass concentrations, like Giant Molecular Clouds (GMC) as natural CR detectors  \citep{aharonianetal,aharonian01,gabici,casanova}. 
The nearest GMCs  form the Gould Belt, a ring-like structure of diameter $\sim 1$~kpc  inclined at $\sim 20^\circ$ to the Galactic Plane  \citep{dame87,perrot03}. CR induced \gr\ emission from the Gould Belt clouds was previously detected by COS-B \citep{caravae,bloemen}, EGRET  \citep{digel,digel99} and by the Large Area Telescope (LAT) on board of Fermi satellite \citep{japanese,fermi_paper}.
  
The bulk of \gr\ emission from the GMCs is produced by GCR interactions. The GCR spectrum may be reconstructed using the measured \gr\ spectrum from the clouds combined with the known relevant particle physics of pion production and decay  \citep{kamae06,kelner}\footnote{A comparison of different codes for calculation of pion production in $pp$ interactions was recently done in Ref. \cite{cholis11}}.  In what follows we report such a measurement. 
 
\begin{table}
\begin{tabular}{|l|c|c|c|c|c|}
\hline
Name& ($l_{\rm  s}, b_{\rm  s})$ & $\Theta_{\rm s}$& $(l_{\rm b},b_{\rm b}$) & $D$& $M$\\
\hline
Perseus OB2& (159.28,-20.22)  &4.0   & (148.43,-19.91)& 350&1.3\\
Taurus            & (173.17,-14.70)   &6.0  & (185.96,-16.86)&140&0.3  \\
Orion A           & (212.23,-19.10)    &4.0  & (222.51,-23.47)&500&1.6 \\
Orion B           & (204.79,-14.15)     &4.0   & (220.92,-22.29)&500&1.7  \\
Mon R2        &   (213.79,-12.58)    &1.5&    (220.92,-22.29)&830&1.2  \\
Chameleon   & (300.42,-16.09) &5.5 & (283.78,-15.88)&215&0.1 \\
Rho Oph        & (355.80,16.63) &5.0 & (4.73,15.90)&165&0.3 \\
 R CrA          & (0.60,-19.64)   &3.0  & (6.12,-22.35)&150&0.03   \\
Cepheus        &(108.54,14.78) & 6.0  & (92.10,13.34)&450&1.9  \\
\hline
\end{tabular}
\caption{High Galactic latitude Gould Belt clouds considered in the analysis.$l_{\rm s}$, $b_{\rm s}$ and $\Theta_{\rm s}$ are the coordinates of the center and radius of regions from which source counts are collected. $l_{\rm b}$, $b_{b}$ are the coordinates of the centers of $5^\circ$ regions used for background estimation. Masses $M$ (in units of $10^5M_\odot$) and distances $D$ (in parsecs) to the clouds are from Ref. \cite{dame87}.}
\vskip-0.5cm
\label{tab:list}
\end{table}

\noindent{\bf Data selection and analysis.}
Clouds from the Gould Belt span large angular sizes $\Theta\sim 1^\circ-10^\circ$. Most of these clouds lie in the Galactic Plane, so that the  \gr\ emission is superimposed on the diffuse Galactic emission. This makes the analysis of  characteristics of cloud \gr\ emission difficult: the diffuse Galactic emission is variable on the angular scales comparable to $\Theta$ and has similar spectral characteristics. Clouds at high Galactic latitudes are separated from the Galactic Plane. Analysis for these clouds can thus be done in a straightforward way.  Taking this into account, we concentrate on the study of high Galactic latitude ($|b|>10^\circ$) clouds listed in Table \ref{tab:list}.

We use the LAT data collected between August 4. 2008 and July 15, 2011. We filter the data with Fermi Science Tools\footnote{http://fermi.gsfc.nasa.gov/ssc/data/analysis/scitools/} (software version {\tt v9r23p1} and  data selection {\tt p6\_v11}) using  {\it gtselect}, {\it gtmktime} and {\it gtbin}. We retain only  \gr\ events ({\tt dataclean} events) at zenith angle $\le 100^\circ$.  We use the aperture photometry  method for the spectral analysis,  by collecting the events from "source" regions and comparing the total number of counts in each source region with the number of background events, estimated from nearby  "background" regions at the same Galactic latitude. The lists of "source" and "background" regions is given in Table  \ref{tab:list}. The exposure is calculated using {\it gtexposure} tool. 

Large regions occupied by the clouds contain point sources. Emission from the point sources is superimposed on the diffuse emission from the clouds.  To subtract the point sources, we use the list of sources from the two year exposure of LAT  \cite{fermi_catalog}. 

The  point spread function (PSF) of LAT becomes comparable to the size of the clouds at energies $\lesssim 0.3$~GeV for photons  pair converted in the "front" layer of the LAT and at $\lesssim 0.7$~GeV for photons converted in the "back" layer.  We use only front photons for the analysis  between 0.3 and 0.7~GeV.  We do not use the $E<300$~MeV data. 

\noindent{\bf Gamma-ray spectrum of the clouds.}
All the clouds listed in Table \ref{tab:list} are detected as extended sources with LAT. The morphology of \gr\ emission  follows the morphology of CO emission as inferred from the CO maps  \citep{dame87,dame01}. The \gr\ flux is proportional to the CO integrated intensity as expected if the flux scales as $F\sim M/D^2$ with $M$ and $D$ being the cloud mass and distance\cite{aharonianetal,aharonian01,gabici,casanova}. Images and spectra of individual clouds can be found in the Supplemental Material.

\begin{figure}
\includegraphics[width=\linewidth]{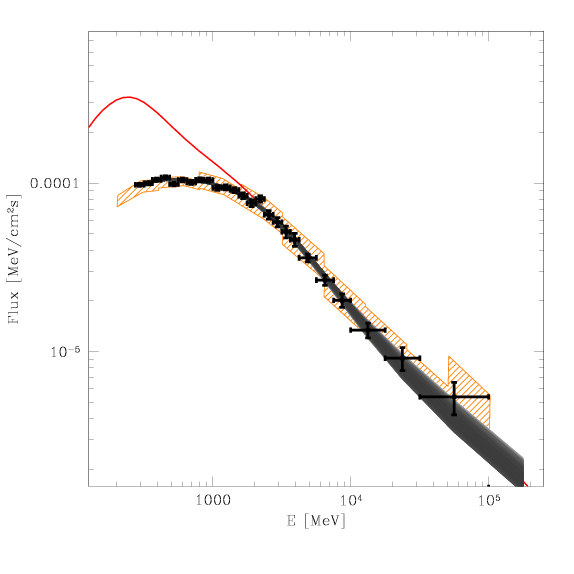}
\caption{Spectrum of diffuse \gr\ emission from high Galactic latitude GMCs. Red thin curve is the spectrum of \gr\ emission  calculated assuming powerlaw CR spectrum. Orange hatched region is "local atomic hydrogen" component from the Ref.  \cite{fermi_background}, renormalized by a factor 23.7 to match the normalization of the GMC average spectrum. Grey shaded area shows model spectra calculated assuming broken powerlaw GCR spectrum.}
\vskip-0.5cm
\label{fig:spectrum_kamae}
\end{figure}

We verified that the spectra of individual clouds  are consistent with each other and with the "local atomic hydrogen" component  of diffuse Galactic emission derived in Ref. \cite{fermi_background} (Fig. \ref{fig:spectrum_kamae}).  This implies that all the high-latitude Gould Belt clouds are "passive" CR detectors, with no on-going particle acceleration and no modification of CR propagation inside the clouds  (i.e. the CR diffusion coefficient does not change significantly in the clouds \cite{gabici}). The average spectrum of the clouds is shown in Fig. \ref{fig:spectrum_kamae}. 

From Fig. \ref{fig:spectrum_kamae} we see that the \gr\ spectrum exhibits a break at $\sim 2$~GeV.  Several explanations of this break  can be considered: a break in the spectrum of CR nuclei or CR electrons/positrons or a break in the \gr\ production cross-section. 

 \gr s with energies $\sim 2$~GeV  are produced by protons with energies much higher than the pion production threshold, where proton-proton interaction cross-section grows logarithmically with energy. The break could also not be related to a feature in the electron Bremsstrahlung cross-section since this also grows logarithmically at GeV 
energies. The most reasonable explanation for the $\sim 2$~GeV feature is that it is related to a feature in the CR spectrum. 

\noindent{\bf CR spectrum.}
We have reconstructed the spectrum of CRs from the \gr\ spectrum using the parametrized pion production spectra calculated in Ref. \cite{kamae06}.  Fig. \ref{fig:spectrum_kamae} shows a comparison  of  the observed \gr\ spectrum with that produced by a powerlaw distribution of CR protons/nuclei. The spectrum produced by a powerlaw CR spectrum has a peak at $\sim 300$~MeV and the model over-predicts the \gr\ emission below $\sim 1$~GeV. The sub-GeV flux could be suppressed if the GCR spectrum hardens below $\sim 10$~GeV. 

To find the details of the low-energy hardening  we model the spectrum as a broken powerlaw   $dN_{CR}/dE\sim\left.\left(E/E_{\rm Br}\right)^{\Gamma_1}\right/\left(1+\left(E/E_{\rm break}\right)^\sigma\right)^{(\Gamma_2+\Gamma_1)/\sigma}$ with low and high energy slopes $\Gamma_1$, $\Gamma_2$, break energy $E_{\rm break}$ and  sharpness of the break $\sigma$.  Assuming negligible Bremsstrahlung contribution, a satisfactory fit to the \gr\  spectrum is found, with $\chi^2/\mbox{d.o.f}=23/21$ which corresponds to a 34\% probability for model to be the proper description of the data. The range of model \gr\ spectra consistent (at 68\% confidence level) with the GMC data is shown in Fig. \ref{fig:spectrum_kamae}. 
\begin{figure}
\includegraphics[width=\linewidth]{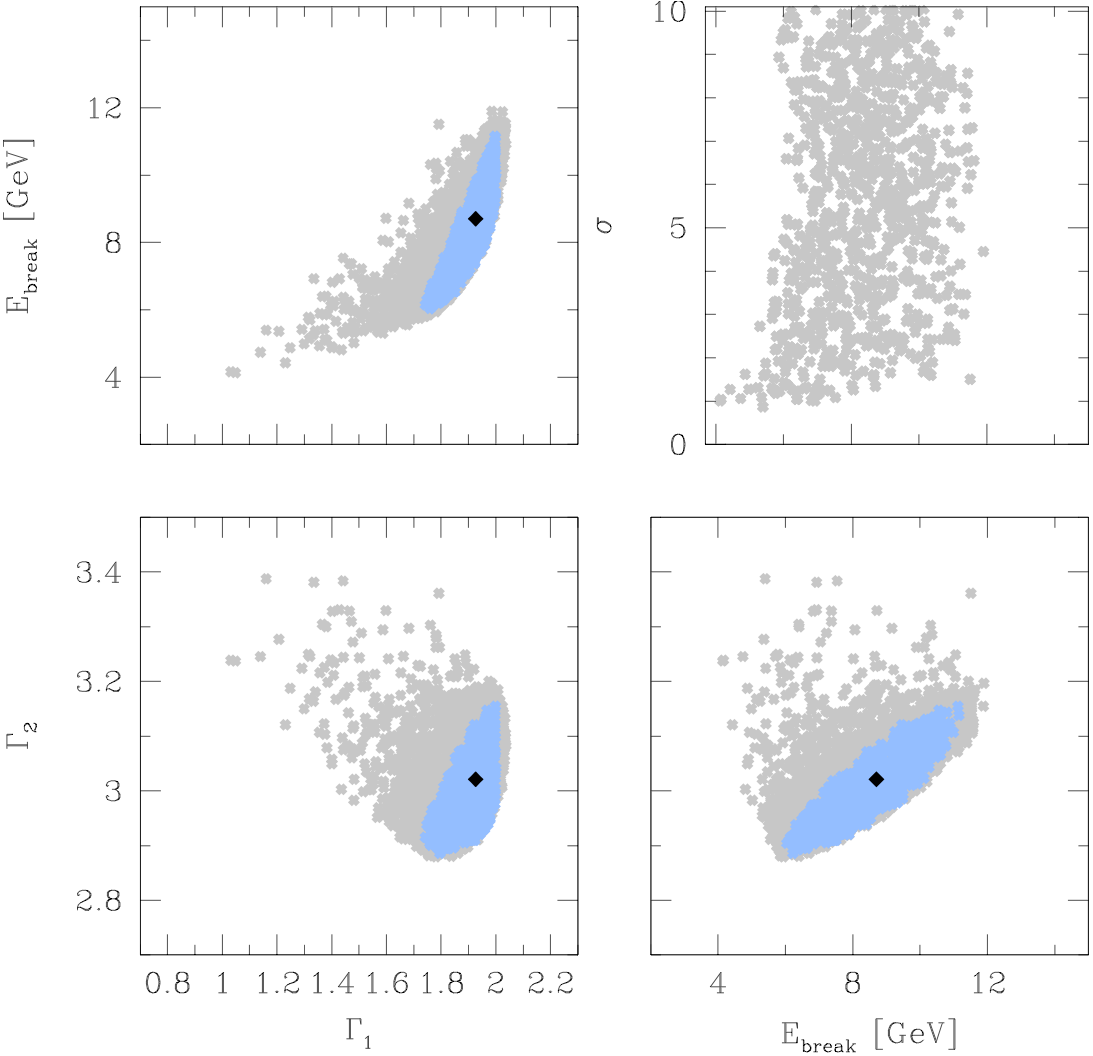}
\caption{68\% confidence ranges for $E_{\rm break}$, $\Gamma_1, \Gamma_2$ and $\sigma$ parameters. Black dot shows the best-fit. Light-blue shaded region is for $\sigma\rightarrow\infty$, grey-shaded region is for finite $\sigma$.}
\vskip-0.5cm
\label{fig:combined}
\end{figure}

The conclusion about the presence of a low-energy break in the CR spectrum is not altered if a non-negligible Bremsstrahlung component is considered. The maximal Bremsstrahlung component is produced when electrons  (primary CR electrons and secondary electrons produced in interactions of CR nuclei) loose all their energy via Bremsstrahlung before leaving the clouds. This could happen if electron diffusion through the clouds is slower than in the ISM. This is not realized in the Gould Belt, but considering the maximal Bremsstrahlung component allows an estimate of influence of Bremsstrahlung-related uncertainties.  Including the maximal Bremsstrahlung component we  find a satisfactory fit  ($\chi^2/\mbox{d.o.f}\simeq 24/21$) which is achieved with a sharper break in the CR spectrum, because  Bremsstrahlung contributes mostly to the lower energy part of the spectrum at $E\lesssim 300$~MeV \cite{gabici}.
 
To obtain the estimates for $E_{\rm break}$, $\Gamma_1$, $\Gamma_2$, $\sigma$ we follow Ref. \cite{lampton76}.  Projections of the 68\% confidence intervals for the model parameters onto 
 $(E_{\rm break},\sigma)$, $(E_{\rm break},\Gamma_2)$, $(\Gamma_1,\Gamma_2)$ and $E_{\rm break},\sigma$ planes are shown in Fig. \ref{fig:combined}.  
 The sharpness of the break, $\sigma$, is constrained from below. The best-fit model is the model  of the form
$dN_{CR}/dE\sim \left( E^{-\Gamma_1},\ E<E_{Br}\right)\bigcup\left( E^{-\Gamma_2},\ E>E_{Br}\right)$. The confidence regions for model parameters in the "sharp break" model are shown as light-blue shaded regions in Fig.  \ref{fig:combined}. Projecting these regions onto coordinate axis one finds  $E_{\rm break}=9\pm 3$~GeV and $\Gamma_2=3.03\pm 0.17$, $\Gamma_1=1.9\pm 0.2$. Considering the model with maximal Bremsstrahlung component one finds a reduced estimate for $\Gamma_1=1.7\pm 0.2$. Leaving parameter $\sigma$ free  increases the errorbars to  $E_{\rm break}=9^{+3}_{-5}$~GeV, $\Gamma_2=3.03^{+0.37}_{-0.18}$, $\Gamma_1=1.9^{+0.2}_{-0.9}$.

\begin{figure}
\includegraphics[width=\linewidth]{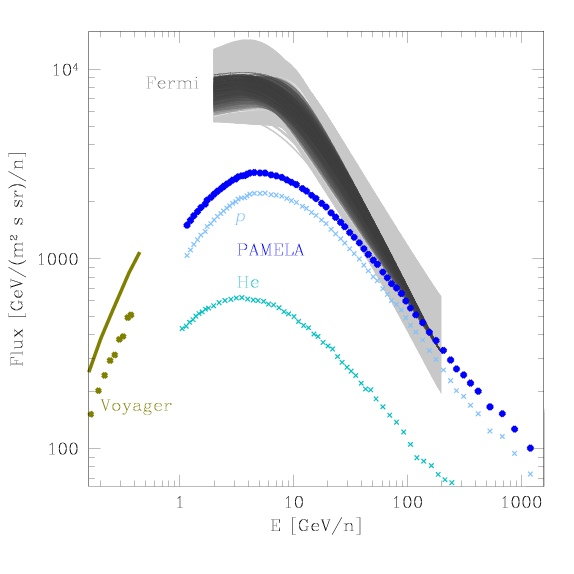}
\caption{ Light grey: GCR spectrum inferred from the LAT observations. Dark grey: GCR spectrum inferred from the LAT observations with  normalization fixed  at 200 GeV to PAMELA.  Thin data points are proton (light blue), helium (cyan) and the total (blue)  CR spectrum measured by PAMELA \cite{pamela}.  Olive data points in 0.1-0.35 GeV range are Voyager data for CR flux beyond the Solar wind termination shock \cite{webber}. Solid curve is GCR spectrum reconstructed from  Voyager \cite{webber,florinski}.}
\vskip-0.5cm
\label{fig:CR}
\end{figure}

The PAMELA collaboration has reported a measurement of  the CR spectrum  between $50\mbox{ GeV}<E< 200$~GeV with the spectral slope of $\Gamma\simeq 2.85$  \cite{pamela}. LAT measurements are mostly sensitive to the part of the CR spectrum below 200~GeV. The slope of the CR spectrum $\Gamma_2$  is consistent with the PAMELA measurement in the $50-200$~GeV range. Below 50~GeV the CR spectrum derived from Fermi measurement is still consistent with a powerlaw down to $\simeq 10$~GeV, while the spectrum measured by the PAMELA deviates from the powerlaw.

Contrary to the shape, the normalization of the GCR spectrum is more difficult to measure using the \gr\ data. The problem here being the large uncertainty (by a factor of $\sim 2$) of the of amount of  target material in the clouds. To remove this uncertainty, we normalize the CR flux above $200$~GeV in the PAMELA data. This is justified because the effect of the Heliospheric distortion at this energy is minor.  The GCR spectrum normalized in this way is shown by dark shaded curve in Fig. \ref{fig:CR}.

An alternative possibility to find the normalization of GCR spectrum is to infer the  \gr\ emissivity per hydrogen atom which is determined by the CR density.  Such a measurement relies on an estimate of the hydrogen column density from the  CO maps using the CO-to-N$_{H_2}$ conversion "X-factor". Assuming  $X=1.8\times 10^{20}$~cm$^{-2}$K$^{-1}$km$^{-1}$s \cite{dame01} we derive the normalization of the CR spectrum shown by the light grey band in Fig. \ref{fig:CR}. A constant nuclear enhancement factor, $\kappa\sim 1.5 - 1.8$, which accounts for the nuclear composition of the CR flux and of the interstellar medium, was assumed \cite{delahaye11,mori09}. Statistical uncertainty of such measurement ($\sim 10\%$) is much smaller than the systematic uncertainty related to uncertainty of the X-factor \cite{fermi_paper} ($\sim 40\%$) and by the uncertainty of $\kappa$ ($\sim 20$\%) \cite{delahaye11,mori09}). 

\noindent{\bf Discussion.}
The LAT observation of the high Galactic latitude clouds from the Gould Belt shows that the steepening of GCR spectrum below $\sim 200$~GeV persists down to $\sim 10$~GeV.  The PAMELA spectrum below $\sim 50$~GeV deviates from the GCR spectrum derived from the LAT data. This could be attributed to the distortion of the GCR spectrum in the Heliosphere. In the conventional modeling, the Heliosphere is assumed to affect the CR flux only below $\sim 10$~GeV, in the energy band where the Solar modulation, or time variability of the flux is observed \cite{bess}. However, the heliospheric effects might affect the CR spectrum up to TeV energies at which the gyroradius of CRs becomes comparable to the size of the Heliosphere, $\sim 100$~AU. At 50~GeV,  the  gyroradius  is $R_L=E/eB\simeq 1\left[E/50\mbox{ GeV}\right]\left[B/10\ \mu\mbox{G}\right]^{-1}$~AU, comparable to the size of magnetic structures in the outer Heliosphere recently revealed by the Voyager spacecrafts \cite{opher11}. The possibility that the Heliosphere influences the CR flux at the energies higher than $10$~GeV, which though physically justifiable, needs to be further investigated. Residual influence of the Heliosphere might be present up to still higher energies, $\sim 1-10$~TeV, at which the anisotropy of the CR flux in the direction toward and opposite of the Heliotail is observed \cite{amenomori}. 

The GCR spectrum breaks by $\Delta\Gamma\simeq 1$ below $\sim 10$~GeV.  Such a break  was not reported before, although indications for the existence of a break, based on comparison of the CR data with GCR propagation models, were discussed \cite{moskalenko02,cholis11,trotta11}. A possibility of existence of a break in the spectrum of CR electrons in the same energy range was recently discussed \cite{strong11}. The main difference between our result and that of Refs. \cite{moskalenko02,cholis11,trotta11} is that we obtain a direct and model-independent measurement of the break, without any assumptions on the distribution of sources in the Galaxy and details of CR propagation. A potential uncertainty of our measurement  is possible effect of CR propagation in the clouds on the measured spectrum. Absence of cloud-to-cloud variations of the position of the break argues against such a possibility. Small cloud-to cloud variations might be detectable with deeper exposure by LAT.

The hard slope of GCR spectrum below the break, $\Gamma_1\lesssim 2$, is important because it insures a finite energy density of GCRs \cite{gloekler67}. Upper limits on the GCR flux in the 0.1-0.35 GeV range derived from the Voyager measurements \cite{webber,florinski} (Fig. \ref{fig:CR}). Combining the Voyager constraint with the LAT measurements we derive a measurement of the GCR energy density, $U_{CR}=0.9\pm0.3$~eV/cm$^3$. Such an energy density is in equipartition with the energy density of magnetic fields $U_B\simeq 1\left[B/6\ \mu\mbox{G}\right]^{1/2}$~eV/cm$^3$ and turbulent motions $U_{turb}\simeq1\left[n_{ISM}/1\mbox{ cm}^{-3}\right]\left[v_{turb}/20\mbox{ km/s}\right]^2$~eV/cm$^3$ of the ISM with density $n_{ISM}$ and turbulent velocity $v_{turb}$, a fact which points to a physical coupling between the three ISM components.

The detection of a low-energy break introduces a new energy scale into the CR physics. Taking into account that the new energy scale is not far from the "natural" scale of proton rest energy $m_pc^2$, we have tested a possibility that the new scale could be reduced to $m_pc^2$ if the broken powerlaw model of CR spectrum is changed to an alternative model of the form $dN_{\rm CR}/dE\sim \beta^{p_1}{\cal R}^{p_2}$, where $\beta$ and ${\cal R}$ are CR velocity and rigidity. Such shape of GCR spectrum does not have additional scale except for $m_pc^2$. We find that such a GCR spectrum is inconsistent at $>4\sigma$  with the LAT data, with the best-fit value $\chi^2/d.o.f.>3$ for 23 d.o.f.

A change of the CR spectral slope might be related to the physics of CR sources (e.g. characteristic maximal/minimal/break energy of CRs produced by a source population \cite{moskalenko02,cholis11,trotta11}, or injected by annihilation/decay of Dark Matter) or to a change of the propagation for CRs (e.g. transition from convective to diffusive regime \cite{aharonian01,moskalenko02} or change in the energy dependence of the diffusion coefficient \cite{moskalenko02,putze11}  and/or diffusive re-acceleration possibly combined with an intrinsic break in the source spectra \cite{trotta11,putze11}). 

A low-energy cut-off in the source spectra could occur e.g. if the lower-energy particles are trapped by magnetic fields inside the sources. CRs of  energies $\sim 10$~GeV propagate through the Galaxy in a diffusive way by scattering on turbulent inhomogeneities in ISM.  A feature in the turbulence spectrum at a length scale $\lambda_T$ might produce a break in the CR spectrum at an energy at which the Larmor radius $R_L=E_{CR}/eB$ ($B$ is magnetic field in the ISM) is comparable to $\lambda_T$. Scattering and/or absorption of the lower energy CRs would be determined by the energy-independent geometrical cross-section of the smallest ISM inhomogeneities. In such a model, measurement of the break energy $E_{\rm break}\sim 10$~GeV implies the detection of a feature in the distribution of inhomogeneities of ISM at the length scale $\lambda_T\sim E_{\rm break}/eB\sim 1$~AU. 

Suppression of the low energy CR flux might also occur through efficient CRs interactions with the ISM on the time scale of proton-proton interactions $t_{pp}\simeq 3\times 10^7\left[n_{ISM}/1\mbox{ cm}^{-3}\right]^{-1}$~yr. During this time, CRs could spread over a region of the size $R(E)\lesssim \sqrt{D(E)t_{pp}}\sim 1\left[E/1\mbox{ GeV}\right]^{0.3}\left[n_{ISM}/1\mbox{ cm}^{-3}\right]^{-1/2}$~kpc around a CR source (assuming $D(E=1\mbox{ GeV})\sim 10^{28}$~cm$^2$/s for the CR diffusion coefficient). If the distance to the nearest GCR accelerator is in the kiloparsec range, energy losses would efficiently remove CRs with energies below several GeV from the locally observable GCR flux. This mechanism of suppression of the GCR flux could work only  if the last episode of injection of GCR within a ~1 kpc volume occurred not later than $t_{GCR}\sim t_{pp}\sim 3\times 10^7$~yr ago.  Remarkably, this estimate is close to the age of the Gould Belt, which was formed in an explosive event some $t_{GB}\simeq 3\times 10^7$~yr ago \cite{perrot03}.

{\it Acknowledgement}.  This work was supported by the Swiss National Science Foundation grant PP00P2\_123426/1. We are grateful to M.Audard, A.Carmona and F.Aharonian for discussions of the subject.

\newpage
\onecolumngrid
\section*{Supplemental Material}
\subsection*{S1. Imaging analysis}

All the clouds listed in Table 1 are detected as extended sources with LAT. Comparison of the countmaps shown in Fig. S1-S7 for the energy band $E>1$~GeV with the CO intensity [11,13] shown by contours reveals a good correlation between the \gr\ and CO emission. Regions used to estimate the source and background fluxes for each cloud (Table 1) are shown as white solid and dashed circles in Figs. S1-S7.

\vskip0.3cm
\includegraphics[width=0.8\linewidth]{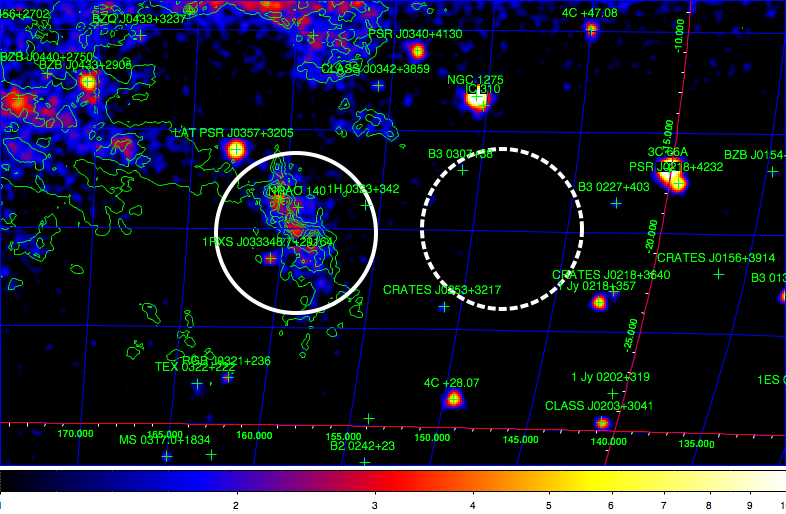}\\
Figure S1. Count map  of Perseus OB2 cloud region in the energy range $E>1$~GeV smoothed with a Gaussian of the width $0.3^\circ$. Green contours show CO emission intensity with the levels 5, 15, 25, 35 and 45 K km/s. Crosses show positions of sources from the two-year Fermi catalog [24]. Solid and dashed white circles are regions used to estimate the source and background fluxes.

\vskip0.3cm
\includegraphics[width=0.8\linewidth]{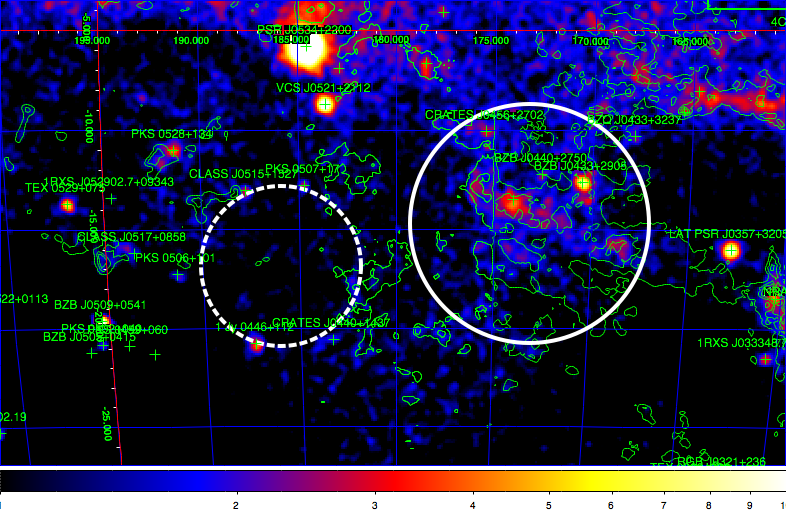}\\
Figure S2. Same as in Fig. S1 but for the Taurus cloud.

\vskip0.3cm
\includegraphics[width=0.8\linewidth]{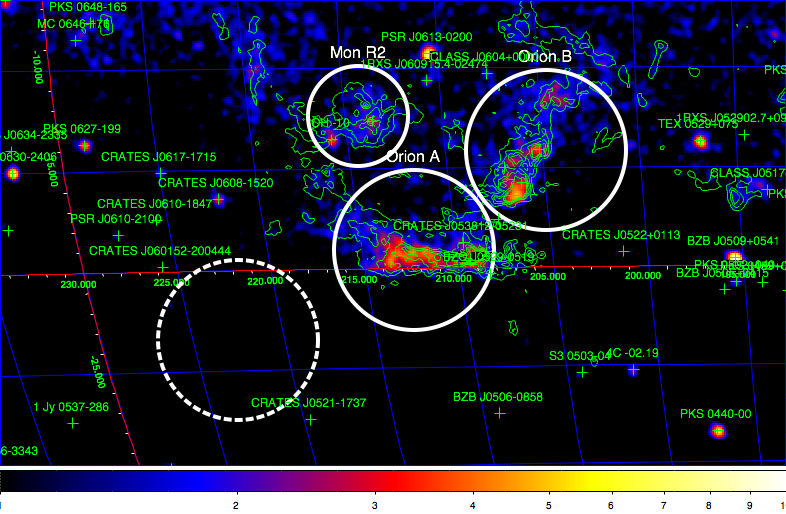}\\
Figure S3. Same as in Fig. S1 but for the Orion cloud region. 

\vskip0.3cm
\includegraphics[width=0.8\linewidth]{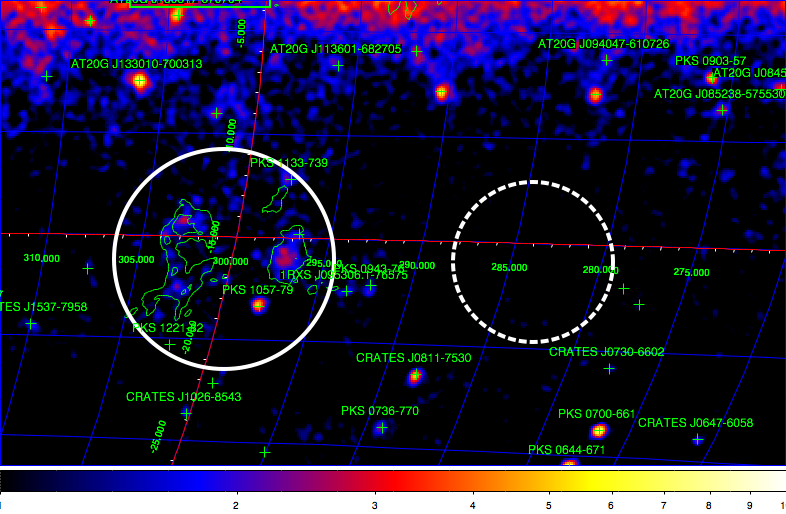}\\
Figure S4. Same as in Fig. S1 but for the Chameleon cloud region. 

\vskip0.3cm
\includegraphics[width=0.8\linewidth]{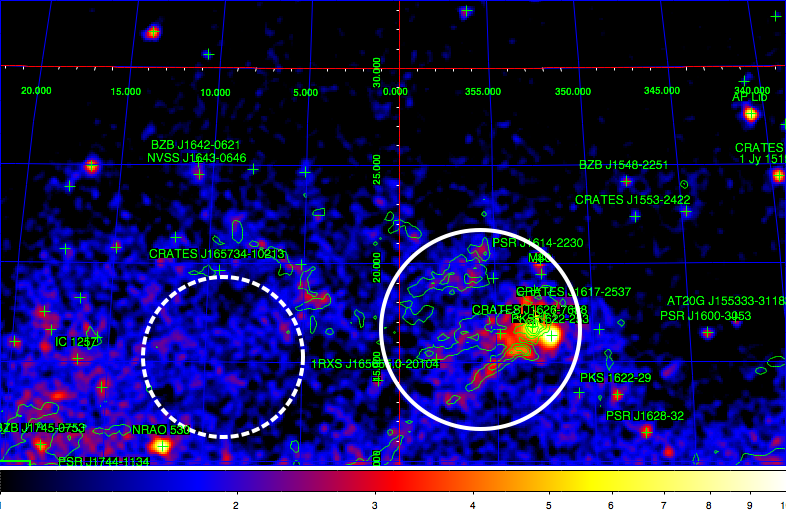}\\
Figure S5. Same as in Fig. S1 but for the $\rho$ Ophiuchus cloud region. 

\vskip0.3cm
\includegraphics[width=0.8\linewidth]{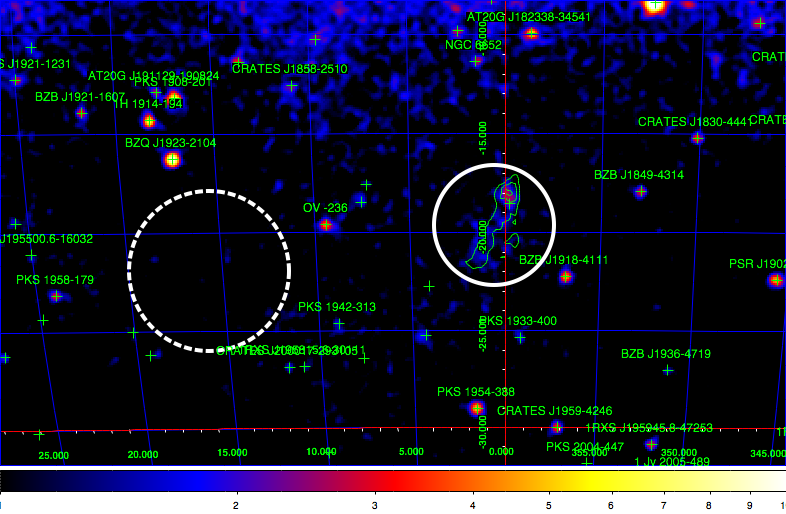}\\
Figure S6. Same as in Fig. S1 but for the r Cr A cloud region.

\vskip0.3cm
\includegraphics[width=0.8\linewidth]{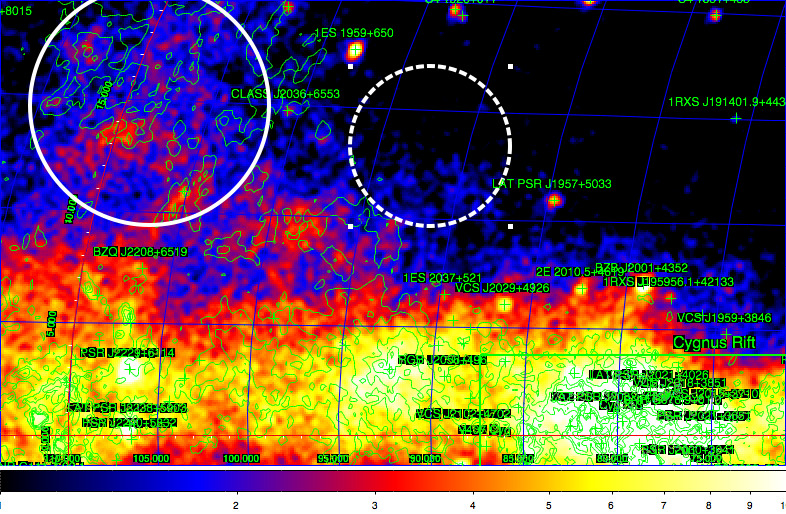}\\
Figure S7. Same as in Fig. S1 but for the Cepheus cloud region. 

\subsection*{S2. Spectral analysis}
\subsubsection*{S2.1 Perseus OB2}

\vskip0.3cm
\includegraphics[width=0.4\linewidth]{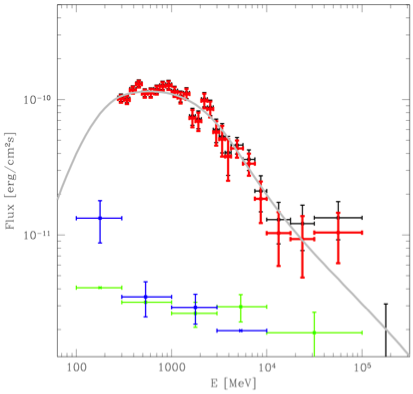}
\includegraphics[width=0.4\linewidth]{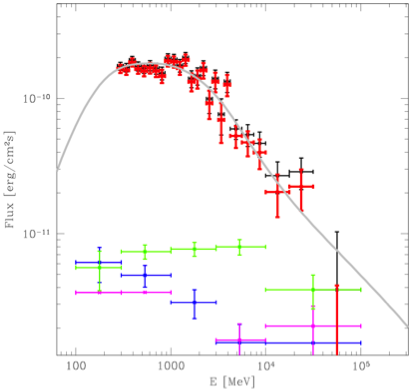}\\
Figure S8. Spectra of Perseus OB2 (left) and Taurus (right)  clouds. Black data points show the total background subtracted spectra of the source regions, red data points are the spectra of diffuse emission. Other data points are spectra of point sources from the two-year Fermi catalog [24] which are subtracted from the total source region spectrum in the estimate of diffuse emission spectrum. Left panel:  blue and green thin data points show the spectra of NRAO 140 and 1RXS J033348.7+29164. Right: green: BZB J0433+2905; blue: CRATES J0456+2702; magenta: BZB J0440+2750. Points without y errorbars are upper limits.
\vskip0.3cm

From Fig. S1 we see that two point sources, identified with a quasar NRAO 140 and an X-ray source 1RXS J033348.7+29164 [24], are situated in the middle of diffuse emission region. Spectra of these sources reported in the two-year Fermi catalog  are shown by the blue and green thin data points in Fig. S8, left. We subtract fluxes of these sources from the overall flux of the source  region to obtain the estimate of diffuse emission from Perseus OB2 cloud (shown in red in Fig. S8, left panel).

\subsubsection*{S2.2 Taurus cloud}

Taurus molecular cloud is closest to the Solar system GMC situated in the center of the Gould Belt [11]. It spans a large angular size of about $15^\circ\times 15^\circ$ in the direction close to the Galactic Anticenter. A large sub-cloud situated in the Auriga sky region is at Galactic latitude $|b|<10^\circ$. We do not consider this part of the cloud in our analysis.

An unidentified source 2FGL J0440.5+2554c is clearly associated to the cloud, being positionally coincident with the excess of CO emission (Fig. S2). We consider this source as a part of diffuse emission from the cloud.

The source region of Taurus cloud, used for spectral extraction, contains two blazars BZB J0433+2905, BZB J0440+2750 with the spectra shown in Fig. S8. Spectra of these blazars are subtracted from the overall spectrum of the region in our spectral extraction procedure.

\subsubsection*{S2.3 Orion A.B and Mon R2 clouds}

The Orion molecular cloud is conventionally divided onto A and B sub-clouds, which are clearly distinguishable in the count map shown in Fig. S3.  

\vskip0.3cm
\includegraphics[width=0.4\linewidth]{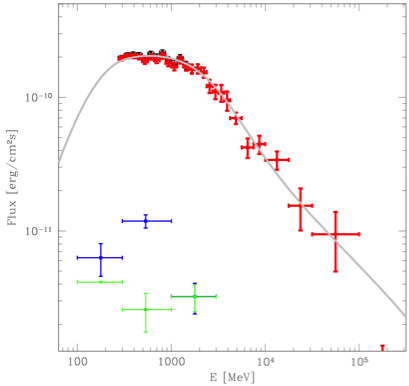}
\includegraphics[width=0.4\linewidth]{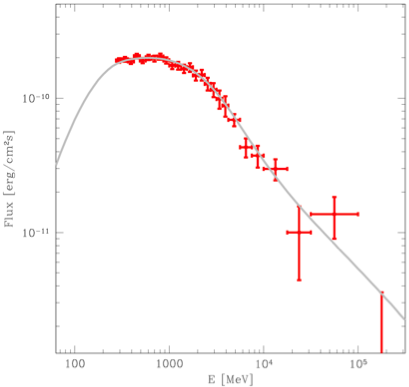}\\
Figure S9. Same as in Fig. S8, but for Orion A (left) and Orion B (right) clouds. Left panel: blue and green are spectra of Crates J0538-05231; green: BZQ J0529-0519. 
\vskip0.3cm

The unidentified Fermi sources 2FGL J0547.1+0020c, 2FGL J0547.5-0141c, 2FGL J0543.2-0120c, 2FGL J0541.8-0203c, 2FGL J0534.9-0450c, 2FGL J0534.8-0548c are most probably associated with emission from the Orion clouds themselves, as can be judged from the correlation of source positions with the excesses in the CO emission (Fig. S3). We do not subtract the spectra of these sources from the overall emission spectrum from the Orion cloud source regions when estimating the spectrum of diffuse emission. 

Sources BZQ J0529-0519 and CRATES J053812-05231 situated inside the region used to estimate the source flux are identified with Flat Spectrum Radio Quasars (FSRQ) in the two-year Fermi catalog [24] are subtracted from the overall source region spectrum (Fig. S9).

Mon R2 cloud region contains a point source OH-10, which is clearly not associated to the cloud emission and is subtracted in the spectral extraction procedure (Fig. S10). The cloud has small angular extent ($2^\circ$ radius circle is used in the spectral extraction). Taking this into account we consider only the data at the energies where the LAT PSF becomes smaller than the size of the region used for the spectral extraction and disregard the data below 1 GeV for this cloud (Fig. S10).

\subsubsection*{S2.4 Chameleon cloud}

\vskip0.3cm
\includegraphics[width=0.4\linewidth]{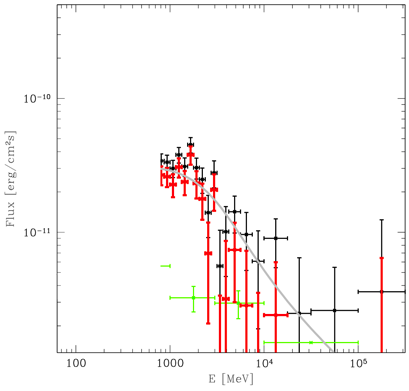}
\includegraphics[width=0.4\linewidth]{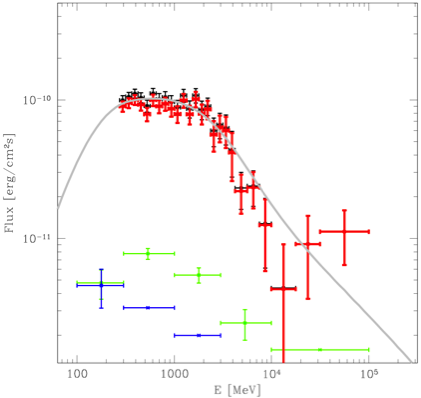}\\
Figure S10. Same as in Fig. S8 but for  Mon R2 (left) and Chameleon (right) clouds. Left panel: green data are the spectrum of OH -10. Right:  green data are the spectrum of PKS 1057-79,  blue data are for  PKS 1221-82. 
\vskip0.3cm

Contrary to Perseus OB2, Orion A, B and Mon R2 clouds, which are situated in a low background Galactic Anticenter region, the Chameleon cloud is situated much closer to the Galactic Center, at Galactic longitude $l\sim 300^\circ$. However, at Galactic latitude $b\sim -15^\circ$ the diffuse \gr\ emission from the cloud region is clearly separated from the Galactic Plane diffuse emission as it is clear from Fig. S4.	

Background sources PKS 1057-79, PKS 1221-82 and PKS 1133-739 are clearly not associated to the cloud. We subtract the spectra of these sources from the overall source region emission when estimating the spectrum of diffuse emission from the cloud (Fig. S10).

\subsubsection*{S2.5 Rho Ophiuchus cloud}

$\rho$ Ophiuchus  cloud is situated in the Galactic Center region with strongly variable Galactic diffuse emission extending to high Galactic latitudes and numerous point sources (Fig. S5).  

A pulsar PSR J1614-2230 could be clearly identified as a point source unrelated to the cloud diffuse emission. PKS 1622-253, a source identified with a blazar, not associated to an excess of CO emission, is, most probably, also a background source. However, the spectrum of this source, shown in Fig. S11, is similar to the spectrum of diffuse emission from the Gould Belt clouds. This might indicate that the source spectrum reported in the two-year Fermi catalog [24] is contaminated by the diffuse emission from $\rho$ Ophiuchus cloud. In this case subtracting the source spectrum from the overall spectrum extracted from the $\rho$ Ophiuchus region we slightly under-estimate the diffuse flux from the $\rho$ Ophiuchus cloud by up to 20\%.

Situation is more complicated with CRATES J1626-7638. This source is identified in the two-year Fermi catalog as a blazar. At the same time, it is spatially coincident with a strong excesses in the CO map. Moreover, the source spectrum repeats the spectrum of diffuse emission from the Gould Belt clouds (Fig. S11). Taking this into account this uncertainty, we do not subtract the spectrum of CRATES J1626-7638 from the overall spectrum of $\rho$ Ophiuchus source region when extracting the spectrum of diffuse emission from the cloud.

\vskip0.3cm
\includegraphics[width=0.4\linewidth]{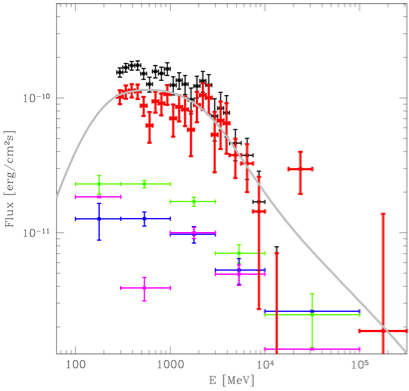}
\includegraphics[width=0.4\linewidth]{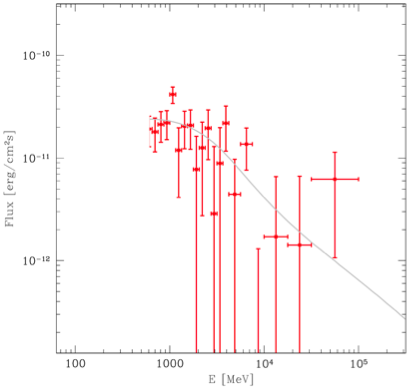}\\
Figure S11. Left: spectrum of  $\rho$ Ophiuchus cloud (red). Green: spectrum of PKS 1622-253; magenta: spectrum of PSR J1614-2230;  blue: spectrum of CRATES J1626-7638. Right: spectrum of r Cr A cloud. 
\vskip0.3cm

\subsubsection*{S2.6 Corona Australis cloud}

Corona Australis (r Cr A) cloud is located in the Galactic Center region at Galactic latitude $b\simeq -20^\circ$. In spite of the low mass of the cloud, \gr\ emission is detectable because of the proximity of the source. 

Maximum of the \gr\ signal coincides with an excess in CO emission (Fig. S6) and is located at the position of an unidentified Fermi source 2FGL J1904.9-3720c. We propose that this source is associated with the cloud emission and do not subtract the spectrum of this source from the overall spectrum of the source from the total source region spectrum (Fig. S11, right).

\subsubsection*{S2.7 Cepheus cloud}

\vskip0.3cm
\includegraphics[width=0.4\linewidth]{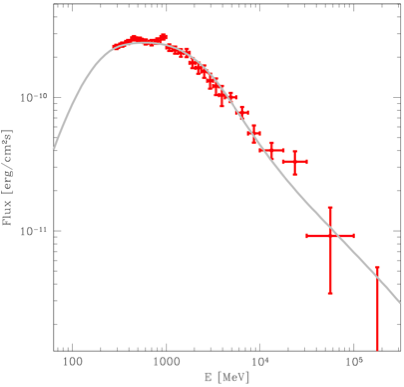}\\
Figure S12. Spectrum of Cepheus cloud (red). Grey: average cloud spectrum.
\vskip0.3cm

Cepheus cloud is a bright and very extended \gr\ source situated just above the Galactic Plane in vicinity of Cygnus regon (Fig. S7). An unidentified source from the two-year Fermi catalog [24] is clearly associated to a strong excess in the CO emission map and we propose that this source should be identified as a clump in  the cloud. We do not subtract the spectrum of this point source from the in our estimate of diffuse emission from Cepheus cloud (Fig. S12).

\end{document}